# The State of Open Access in Germany: An Analysis of the Publication Output of German Universities

Neda Abediyarandi and Philipp Mayr

*neda67a@gmail.com; philipp.mayr@gesis.org*
GESIS - Leibniz Institute for the Social Sciences, Cologne, Germany

**Introduction**

Starting with the Berlin declaration in 2003, Open Access (OA) publishing has established a new era of scholarly communication due to the unrestricted electronic access to peer reviewed publications. OA offers a number of benefits like e.g. increased citation counts (Gargouri et al., 2010) and enhanced visibility and accessibility of research output (Tennant et al., 2016). The OA movement with its powerful mandating and policymaking has been very successful in recent years. Relatively little is known about the real effects of these activities in terms of OA publication output of institutions on a larger scale (Piwowar et al., 2018). The aim of this article is to investigate to what extent the OA fraction of the publication output of German universities has increased in the last years. To answer this question, we analysed and compared total number of publications which have been published by researchers of the largest German universities. We compared the numbers of OA versus closed publications for 66 large German universities in the time span of 2000-2017.

**Methodology**

We follow the classic definitions and classify publications into three categories: Green OA, Gold OA and Closed. Closed access journals allow papers to be read by users with a subscription to the journal (Prosser, 2003). There are two major ways for peer reviewed journal articles to OA, publishing in pure OA journals (gold OA) or archiving of article copies or manuscripts at other web locations (green OA) (Björk et al., 2014).

For the analysis we used Web of Science (WoS) and UNPAYWALL[1] (Piwowar et al., 2018) to extract and analyse our data. To identify German university affiliations in WoS, we used data from the Competence Centre for Bibliometrics, in particular the result of the project "Institutional address disambiguation" (Rimmert et al., 2017).

We first selected 66 German universities which have more than 1,900 publications in WoS in a period of 17 years (2000-2017). In the following, we matched all WoS publications of these 66 universities with UNPAYWALL publications. We considered matching based on DOI and title to get precise results. We got round 34% matched publications because a larger number of DOIs for publications in WoS was missing (especially between 2000 and 2002). In the WoS dataset each publication can be affiliated with some authors. To remove redundancy, we randomly allocated each specific publication to one of its authors, in other words, if a publication is written by several authors from different universities; we counted just for one of them. In Table 1 we list the 10 German universities with the most matched WoS publications from 2000 to 2017.

**Table 1. Total number of matched WoS publications by top 10 German universities (2000-2017).**

| University | Matched WoS articles |
|---|---|
| Heidelberg Univ. | 72,556 |
| LMU Univ. | 67,525 |
| Charité Berlin Univ. | 63,949 |
| Technical Munich Univ | 63,641 |
| Bonn Univ. | 54,671 |
| Nuremberg Univ. | 53,289 |
| Karlsruhe Univ. | 51,266 |
| Hamburg Univ. | 48,880 |
| Freiburg Univ. | 47,574 |
| Technical Dresden Univ. | 47,137 |

**Approach**

In the following, we are investigating the percentage of publications of German universities published in gold, green and closed access. In order to answer this question, we analysed our extracted data in two different aggregations.

1. Comparing the number of publications for the top 10 German universities: We analysed and compared the total number of gold, green and closed access publications for the top 10 German universities (see Table 1 and Figure 1) in terms of matched WoS articles.

---

[1] The UNPAYWALL dataset includes millions of articles in which publications were separated based on their access type (Green, Gold and Closed). https://unpaywall.org/

2. Comparing groups of German universities: We grouped 66 German universities into three different groups based on total number of their published WoS publications from 2000 to 2017.

The different groups of universities are the following:
- *Group 1*: 22 German universities which have published more than 31,000 publications (this includes the top 10 universities from Table 1).
- *Group 2*: 22 German universities which have published more than 12,000 and less than 31,000 publications.
- *Group 3*: 22 German universities which have published more than 1,900 and less than 12,000 publications.

We compared the total number of gold, green and closed access publications which were published by each mentioned group in year 2000, 2010 and 2017 separately (see Figure 2). To verify our analysis, we compared our data with the recent CWTS Leiden Ranking from May 2019[2]. We found a good match between our and the Leiden numbers for the German universities.

**Results**

The total numbers of gold, green and closed access publications for top 10 German universities from 2000 to 2017 are shown in Figure 1. Our findings show that all top 10 German universities still tend to publish most publications within the closed access model. If we compare with Figure 2, we see that the ratio of closed access publications is decreasing, but in the year 2017 still 50% and more of the WoS articles are published in closed access.

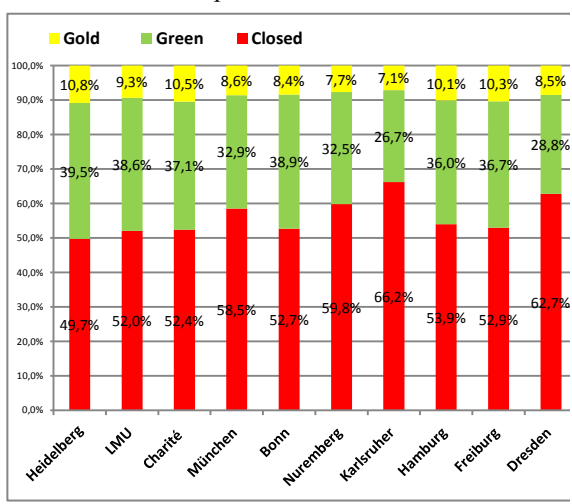

**Figure 1. Comparison of total number of gold, green and closed access publications in the top 10 German universities (2000-2017).**

We found that the top 10 German universities published more gold/green access publications rather than the others. Figure 2 shows the percentage of gold and green access publications for each group are significantly increasing in the last 7 years.

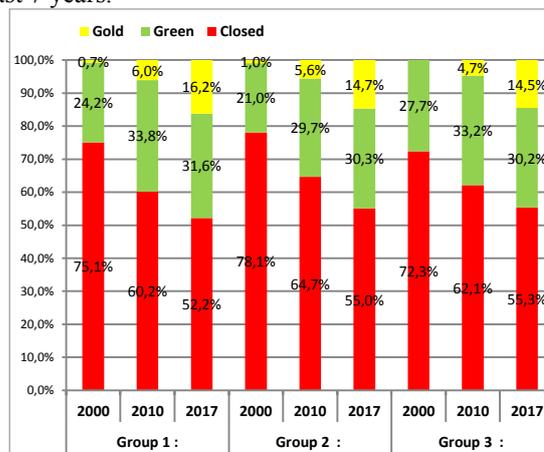

**Figure 2. Three groups of German univ. based on their total number of matched WoS publications (2000, 2010, 2017).**

**Future Work**

As a next step, we plan to analyse the effects of concrete OA mandating in Germany and abroad on the number of green and gold OA publications, their citation advantages and possible enhanced research visibility. In the future, we plan to compare the OA situation in Germany with other European countries and institutions all around the world.

**Acknowledgement**

This work was supported by BMBF project OASE, grant number 01PU17005A.

---

[2] http://www.leidenranking.com/